\documentclass[12pt]{iopart}
\usepackage{iopams}
\usepackage{psfrag,graphicx}
\usepackage{graphicx}
\usepackage{subfigure}
\usepackage[section]{placeins}
\usepackage{psfrag}
\usepackage{caption2}

\begin{document}

\title[Logic gates based all-optical binary half adder...]{Logic gates based all-optical binary half adder using triple core PCF}

\author{T. Uthayakumar$^1$ and R. Vasantha Jayakantha Raja$^{2*}$ }

\address{${^1}$ Department of Physics, Vel Tech Rangarajan Dr.Sagunthala R$\&$D Institute of Science and Technology, Avadi, Chennai 600062,
TN, India\\$^2$ Centre for Nonlinear Science and Engineering, School of Electrical and Electronics Engineering, SASTRA University, Thanjavur 613401, India.\\}
\ead{uthayapu@gmail.com and rvjraja@yahoo.com}

\begin{abstract}
This study presents the implementation of an all-optical binary logic half adder by employing triple core photonic crystal fiber (TPCF). The noteworthy feature
of the present investigation is that an identical set of TPCF schemes, which demonstrated all-optical logic functions in our previous report has revealed the ability
to demonstrate the successful half adder operation. The control signal (CS) power defining the extinction ratios of the output ports for the considered symmetric
planar and triangular TPCFs are evaluated through numerical algorithm. Through suitable CS power and input combinations, the logic outputs are generated from
extinction ratios to demonstrate the half adder operation. The results obtained display the significant influence of the input conditions on the delivery of
half adder operation for different TPCF schemes considered. Furthermore, chloroform filled TPCF structures demonstrated the efficient low power half adder
operation with significant figure of merit, compared to that of the silica counterpart.
\end{abstract}

\maketitle

\section{Introduction}

Nonlinear couplers play an indispensable role in all-optical fiber interconnects as couplers, splitters, wavelength division multiplexors, switches and
optical computers \cite{Agrawal}.  Out of the multitude form of couplers, three core and multiple core couplers attracted special interest for their
manifold output states, good power selectivity with excellent coupling and switching contrast \cite{Crespo,Buah,Silva,Castro,Sarma}.  In recent years,
optical couplers constructed from photonic crystal fibers (PCF) attracted a great deal of attention for their versatile design flexibility and special
optical properties such as high nonlinearity, desired zero dispersion wavelength, high birefringence, large effective mode area, endless single mode
and low bending loss etc \cite{Russell,Hansen,Dudley}. In particular, triple core couplers are shown special attention to realize the long term goal of
achieving an efficient all-optical switching and logic operations. Such all-optical devices, play a significant role in handling large bandwidth as
well as cope up with the speed of the telecommunication interconnects.

A numerous efforts have been made to investigate the performance of photonic crystal fiber couplers (PCFC) for an efficient all-optical functions. Owing
to its significant contribution to the nonlinear and dispersion properties, PCF forms a best choice for the construction of ultrafast all-optical
controlled devices. Based on the coupled mode theory (CMT), the linear and nonlinear characteristics of the optical pulse propagation through triple
core PCF have been analyzed by Li \cite{Li}. Apart from the applications of couplers as switches and splitter, they form a best candidate for the
construction of logic gates based on all optical control. A continuous wave switching is accomplished through three core fiber couplers of planar-
and triangular- configuration with successful demonstration of optical Boolean operations \cite{Menezes1}. Moreover, revealed the various logic
operations and determined the figure of merit numerically for asymmetric two core fiber couplers \cite{Menezes2,Menezes3}. Furthermore a time domain
simulation of an optical AND function in presence of all optical control by employing PCF Y-junction \cite{Danaie} has also been reported.  A successful demonstration of all
all-optical logic gates and its figure of merit for aptly modeled planar and triangular PCFC are realized numerically \cite{Uthayakumar}. Moreover, the
indispensable role of structural asymmetry of PCFC for diverse combinations on the delivery of all optical logic gates has also been reported \cite{Uthayakumar2}.

\begin{figure}[tbp]
\begin{center}
\subfigure[]{\label{f1a}
\begin{minipage}
[b]{0.475\textwidth}
\includegraphics[width=1\textwidth]{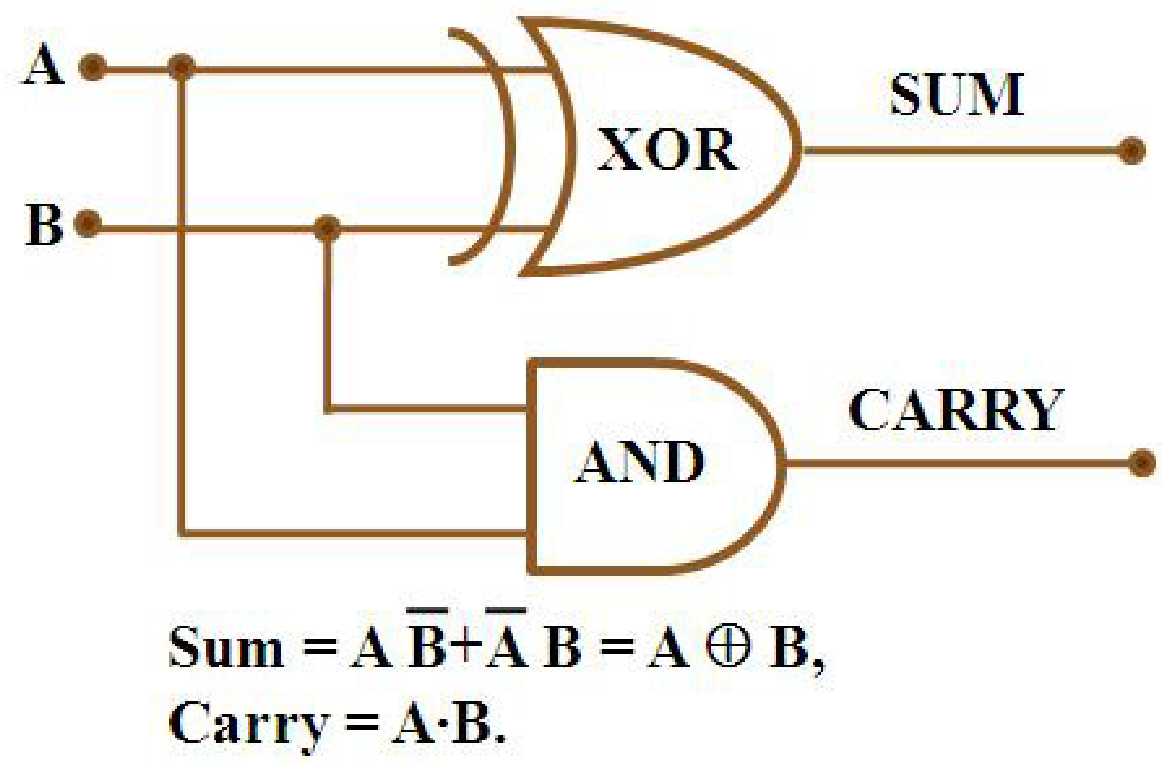}
\end{minipage}}
\subfigure[]{\label{f1a}
\begin{minipage}
[b]{0.475\textwidth}
\includegraphics[width=0.9\textwidth]{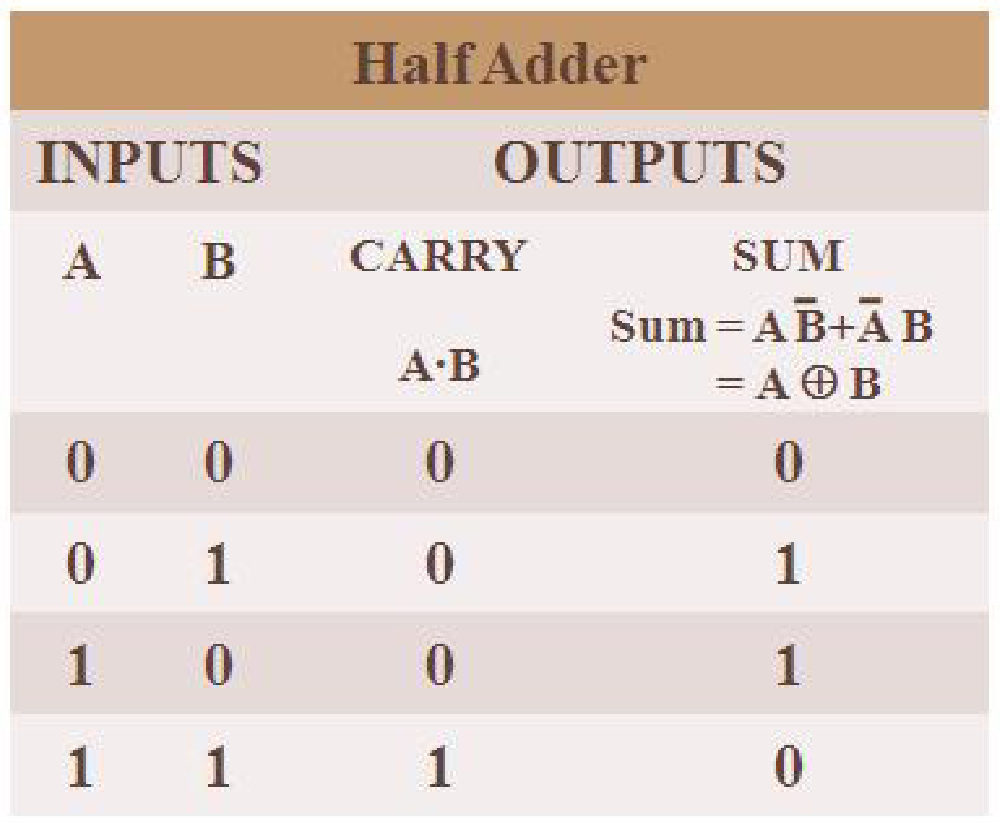}
\end{minipage}}
\end{center}
\caption{ (a) Schematic diagram and (b) Truth table for half adder.}
\end{figure}

Recent reports have proved that the logical functions will allow the realization of devices such as binary counter, half adder and shift
registers \cite{Benner,Poutie,Hall}. Out of these devices, half adder whose function can be achieved through the combination of XOR and AND logic gates form
heart of digital processors. A triple core fiber coupler based numerical study for employing planar coupler to achieve half adder with apt control signal (CS)
has been realized by Menezes et al \cite{Menezes4,Menezes5}.  A schematic and truth table for logic operation of half adder is shown by Fig. 1 (a $\&$ b).  But
till date, the achievement of half adder employing the triple core PCF (TPCF) has not been realized. However, the TPCF with excellent combination nonlinear and
dispersion characteristics allow it to be employed as a best candidate for the digital all optical controlled system to achieve half adder. Hence, the study
intends to explore the transmission characteristics of the TPCF with suitable control signal for the realization of Boolean operation of all optical half adder
through split step Fourier algorithm (SSFM).

The paper is outlined as follows: The TPCF schemes considered and the necessary optical parameters required are provided in the section 2. A numerical study of
the pulse propagation through TPCF described by CNLSE is obtained through the SSFM has been discussed by section 3. The section 4 deals with the construction
of half adder and the verification of the truth table. Finally, the section 5 concludes the paper.

\section{Design of TPCF}

For the proposed study, a similar set of TPCF designs, one with silica as core material (STPCF) and the other with liquid chloroform as core material (CTPCF)
which have proposed in our earlier is considered \cite{Uthayakumar}. The schematic of the considered linear and triangular TPCF configuration are portrayed in
Fig. 2 (a $\&$ b) respectively. The geometrical parameters of the designs considered are same as \cite{Uthayakumar}, namely, air hole diameter ($d$) to the pitch
constant ($\Lambda$) ratio $d$/$\Lambda=$ 0.666, inter core separation ($C\,=2\Lambda$) at $\Lambda=2$ $\mu$m. For the CTPCF, the core diameter ($D_c$) is equal
to that of the air hole diameter ($D_c$ = d). In case of STPCF, the liquid filled cores are removed and the solid silica back ground will serve as guiding cores.
In Fig. 2 (a $\&$ b), the light guiding cores are indicated by 1, 2 and 3. The necessary optical properties for half adder operation are provided by employing
finite element method \cite{Uthayakumara,Uthayakumarb,Uthayakumar,Uthayakumarc,Uthayakumar2}. The planar TPCF configuration is studied for two cases, case (i) planar configuration 1 (PC1) where the input of the
half adder is provided through core 2 and core 3 and control signal will be provided through core 1. Case (ii) planar configuration 2 (PC2) the input is provided
through core 1 $\&$ 3 and the control signal will be provided through core 2. For triangular configuration (TC), the inputs and control signal will be provided
same as that of planar configuration.
\begin{figure}[htb]
\begin{center}
\includegraphics[width=15cm]{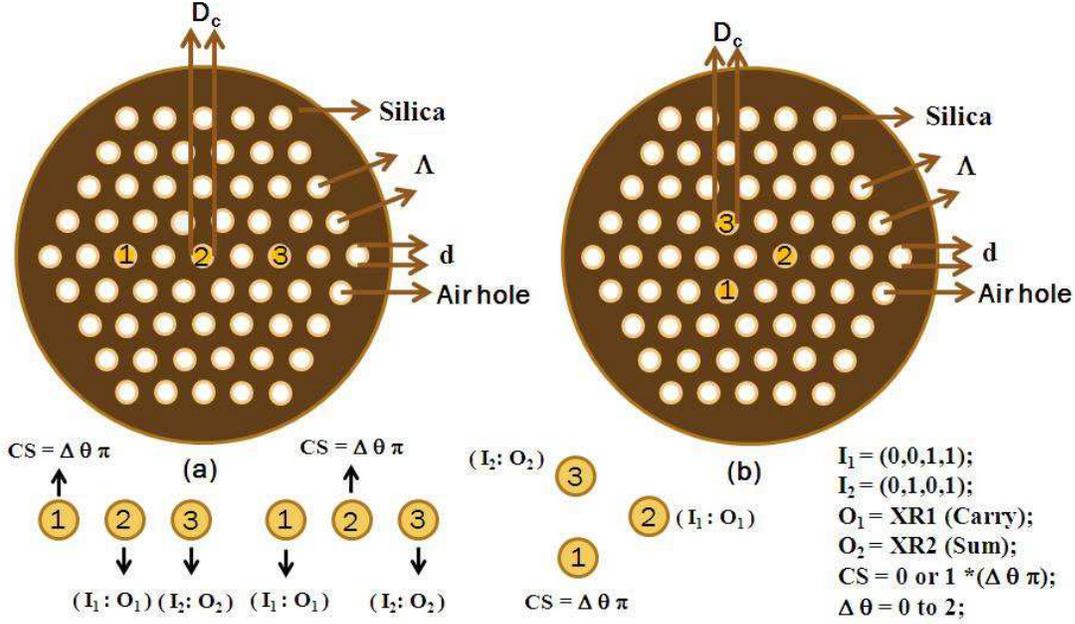}
\end{center}
\caption{Schematic and input configurations of (a) Planar and (b) Triangular TPCF respectively with $\Lambda=2\,\mu m$, $d/\Lambda=0.666$ and $C=2\Lambda$ for half adder operation \cite{Uthayakumar}.}
\end{figure}

\section{Theoretical Model}
\label{Theory}
Optical pulse propagating through the considered TPCF structures are governed by a set of coupled nonlinear Schr\"{o}dinger equations (CNLSE) of the following form \cite{Agrawal,Silva,Uthayakumar}
\begin{eqnarray}
\label{T1}
i\frac{\partial A_1}{\partial z}-\frac{\beta_2}{2}\frac{\partial^2 A_1}{\partial t^2} +\gamma|A_1|^2A_1+ \kappa\,(A_2\,+A_3)-\frac{i}{2}\, \alpha\, A_1=0,\\
\label{T2}
i\frac{\partial A_2}{\partial z}-\frac{\beta_2}{2}\frac{\partial^2 A_2}{\partial t^2} +\gamma|A_2|^2A_2+ \kappa\,(A_1\,+A_3)-\frac{i}{2}\, \alpha\, A_2=0,\\
\label{T3}
i\frac{\partial A_3}{\partial z}-\frac{\beta_2}{2}\frac{\partial^2 A_3}{\partial t^2} +\gamma|A_3|^2A_3+ \kappa\,(A_1\,+A_2)-\frac{i}{2}\, \alpha\, A_3=0,
\end{eqnarray}
Here $A_1$, $A_2$ and $A_3$ describes the input optical pulse propagating through the core 1, core 2 and core 3 respectively. $\beta_2$, $\gamma$  and $\alpha$
represents group velocity dispersion, nonlinear Kerr coefficient and extinction coefficient respectively. The coupling coefficient $\kappa$ is defined by
$\frac{\pi}{2 L_c}$, where $L_c$ is the coupling length. For planar configurations PC1 and PC2, $A_3 = 0$ in Eq. (\ref{T1}) and $A_1 = 0$ in Eq. (\ref{T3}), since there is no interaction between the guiding cores 1 and 3.
For the dynamical study and the calculation of control power from the transmission characteristics of the proposed configurations, refer \cite{Uthayakumar}.

\section{Construction of Half adder}
For the accomplishment of the Half adder, the planar configuration is studied for the two cases (i) For  PC1, control signal (CS) is provided through the core 1 and
(ii) in case of PC2, CS is introduced through the core 2.  The input 1 (I$_1$) and input 2 (I$_2$) are provided through core 2 and core 3 for PC1 and that for PC2 is provided through
provided through core 1 and core 3 for PC2, respectively. And that for the triangular configuration (TC), the CS is provided through the core 1 and I$_1$ and I$_2$ are introduced through core 2 and
core 3. The input, output and CS combinations for half adder operation for all the considered configurations are provided in the bottom of Fig. 2.
The CS applied to all the configurations, may take value 1 or 0 with a phase difference $\Delta\Phi = \Delta\theta\pi$ between the input values I$_1$ and I$_2$
depending upon the logic gates needed. $\Delta\theta$ takes the value from 0 to 2. The output of the logic gates from the core 1 (O$_1$) and core 2 (O$_2$) are
calculated from the extinction ratio XR1 (O$_1$) and XR2 (O$_2$) \cite{Menezes1,Uthayakumar,Uthayakumar2}. By choosing the suitable combinations of I$_1$ and I$_2$ as [(0;0),(0;1),(1;0),(1;1))] and their respective combinations of output (O$_1$) and (O$_2$) are used as a
logic to construct the logic half adder. We assume the initial pulse $A(0,t)=A_0\exp(\frac{-t^2}{W_0})\times\exp(i\Delta\Phi)$ for CS and that of other cores are
either $B(0,t)=A_0\exp(\frac{-t^2}{W_0})$ or 0.001 $\%$ of B(0,t) chosen accordingly to perform the half adder operation.

\subsection{Plane1 configuration}
The XRs of the PC1 for STPCF and CTPCF are shown by Fig. 3. The extinction ratios are same as that of previous configuration but the system can demonstrate only
for certain CS values. For the condition, CS = I$_1$ = I$_2$ = 0, phase values does not influence the logic operation. In case of CS = 0, the output exists only
for the following combinations of inputs [I$_1$;I$_2$]=[(0;1),(1;0),(1;1)] but for CS = 1, output exists for all the input combinations
[I$_1$;I$_2$] = [(0;0),(0;1),(1;0),(1;1)] and obeys for all the configurations considered. From Figs. 3(a) $\&$ 3(b), with CS = 1 ($\Delta\Phi$), I$_1$ = 0 and
I$_2$ = 0, for any phase, the extinction ratio of O$_1$ is around 10 dB, implies that most of the CS power is equally distributed to the O$_1$ and O$_2$. For input
[I$_1$;I$_2$] = [(0;1)], the XR1 value is around -80 dB, which implies that most of input power through I$_2$  and CS power are transferred to the O$_2$ port. In the
case [I$_1$;I$_2$] = [(1;0)], initially a small amount of input power is transferred equally to the output ports O$_1$ and O$_2$ with a minor fluctuations in the
power ratios at the points of greater phase. And the most of output power exits through CS port. Finally, for the case [I$_1$;I$_2$] = [(1;1)], the power variation
exhibits similar behavior as that of the condition [I$_1$;I$_2$] = [(1;0)] and the power level oscillating between the output ports O$_1$ and O$_2$ is further reduced.

\begin{figure}[tbp]
\begin{center}
\subfigure[]{\label{f3a}
\begin{minipage}
[b]{0.475\textwidth}
\includegraphics[width=1\textwidth]{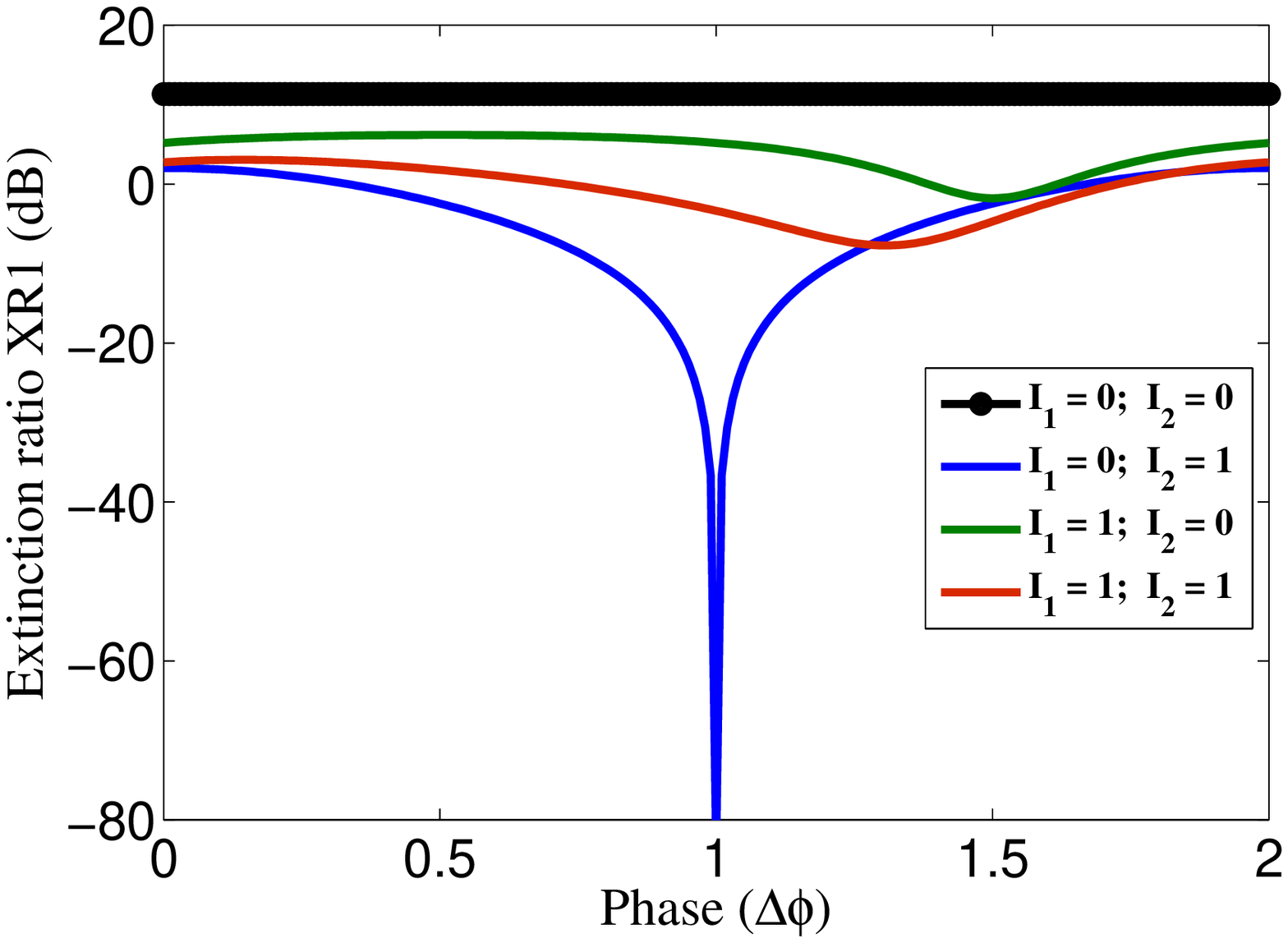}
\end{minipage}}
\subfigure[]{\label{f3b}
\begin{minipage}
[b]{0.475\textwidth}
\includegraphics[width=1\textwidth]{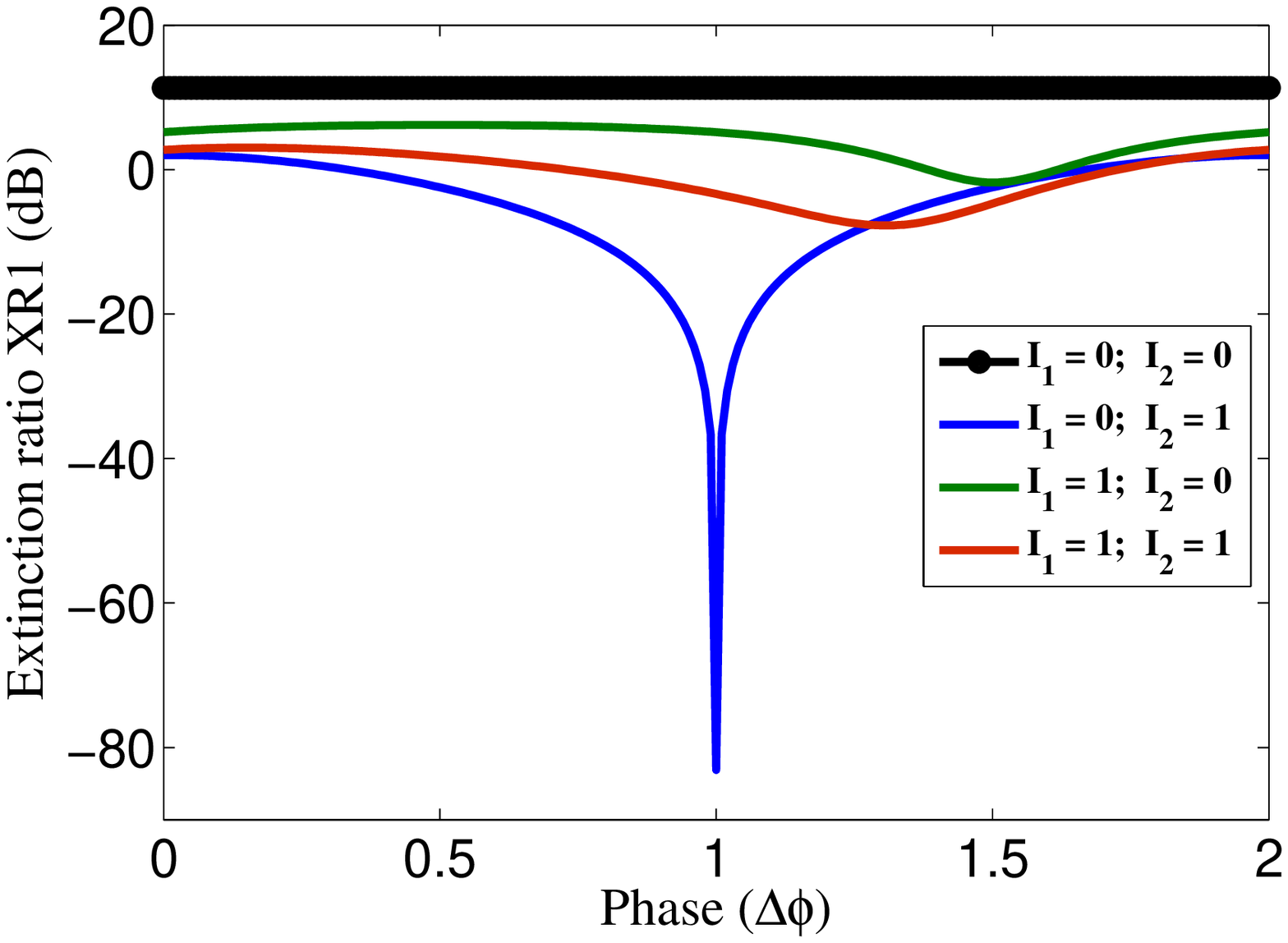}
\end{minipage}}
\end{center}
\caption{ Extinction ratio (XR1) of (a) STPCF and (b) CTPCF of PC1 as a function of the phase parameter \cite{Uthayakumar}.}
\end{figure}

For the half adder function, the SUM and CARRY of the half adder are provided by the outputs O$_1$ and O$_2$ as AND and XOR logic gates respectively. For PC1,
both STPCF and CTPCF demonstrates the half adder operations, only for the phase values $\Delta\theta$ = 1.4, 1.5 and 1.6. The figure of merit (FOM) of half adder
used to identify the best configuration among the half adders achieved for different phase values are defined by the sum of the modulus of the individual
combinations of the extinction ratios to obtain the logic half adder. The maximum figure of merit is achieved for the phase $\Delta\Phi_B$ = 1.4 $\pi$.
Half adder achieved for STPCF and CTPCF with best figure of merit are shown in the Table. 1. Table. 2 provides the FOM calculated for PC1 for all
phase values.

\begin{table}
\caption{Plane 1 configuration (PC1)}
\renewcommand{\arraystretch}{1}
\addtolength{\tabcolsep}{-5pt}
\begin{tabular}{|l|l|l|l|l|l|l|l|}
\hline
\multicolumn{5}{|l|}{\,\,\,\,\,\,\,\,\,\,\,\,\,\,\,\,\,\,\,\,\,\,\,\,\,\,\,\,\,\textbf{Silica}\,\,CS\,($\Delta\phi_B$ = $1.4\,\pi$)}&\multicolumn{3}{l|}{\,\,\,\,\,\,\,\textbf{Chloroform}\,\,CS\,($\Delta\phi_B$ = $1.4\,\pi$)}\\
\hline
\,\,\textbf{I1}\,\,&\,\,\textbf{I2}\,\,&\,\,\,\,\textbf{CS}&\,\,\textbf{O1\,(Carry)}&\,\,\textbf{O2\,(Sum)}\,&\,\,\,\,\textbf{CS}&\,\textbf{O1\,(Carry)}&\,\,\textbf{O2\,(Sum)}\,\\
\hline
\,\,\,0&\,\,\,0&\,\,\,\,\,\,\,0&\,\,\,\,\,\,\,\,\,\,\,\,\,\,\,\textbf{0}&\,\,\,\,\,\,\,\,\,\,\,\,\,\,\textbf{0}&\,\,\,\,\,\,\,0&\,\,\,\,\,\,\,\,\,\,\,\,\,\,\,\textbf{0}&\,\,\,\,\,\,\,\,\,\,\,\,\,\,\textbf{0}\\
\hline
\,\,\,0&\,\,\,1&\,\,\,\,\,\,\,1&\,\,\,\,\,-4.03 dB&\,\,\,\,\,\,4.03 dB&\,\,\,\,\,\,\,1&\,\,\,\,\,\,-4.42 dB&\,\,\,\,\,\,4.42 dB\\
&&&\,\,\,\,\,\,\,\,\,\,\,\,\,\,\,\textbf{0}&\,\,\,\,\,\,\,\,\,\,\,\,\,\,\textbf{1}&&\,\,\,\,\,\,\,\,\,\,\,\,\,\,\,\textbf{0}&\,\,\,\,\,\,\,\,\,\,\,\,\,\,\textbf{1}\\
\hline
\,\,\,1&\,\,\,0&\,\,\,\,\,\,\,1&\,\,\,\,\,-1.49 dB&\,\,\,\,\,\,1.49 dB&\,\,\,\,\,\,\,1&\,\,\,\,\,\,-0.39 dB&\,\,\,\,\,\,0.39 dB\\
&&&\,\,\,\,\,\,\,\,\,\,\,\,\,\,\,\textbf{0}&\,\,\,\,\,\,\,\,\,\,\,\,\,\,\textbf{1}&&\,\,\,\,\,\,\,\,\,\,\,\,\,\,\,\textbf{0}&\,\,\,\,\,\,\,\,\,\,\,\,\,\,\textbf{1}\\
\hline
\,\,\,1&\,\,\,1&\,\,\,\,\,\,\,0&\,\,\,\,\,\,\,2.72 dB&\,\,\,\,\,-2.72 dB&\,\,\,\,\,\,\,0&\,\,\,\,\,\,\,\,2.72 dB&\,\,\,\,\,-2.72 dB\\
&&&\,\,\,\,\,\,\,\,\,\,\,\,\,\,\,\textbf{1}&\,\,\,\,\,\,\,\,\,\,\,\,\,\,\textbf{0}&&\,\,\,\,\,\,\,\,\,\,\,\,\,\,\,\textbf{1}&\,\,\,\,\,\,\,\,\,\,\,\,\,\,\textbf{0}\\
\hline
\multicolumn{2}{l|}{}&\,\textbf{FOM}\,&\,\,\,\,\,\,\,8.25 dB&\,\,\,\,\,\,8.25 dB&\,\textbf{FOM}\,&\,\,\,\,\,\,\,\,7.53 dB&\,\,\,\,\,\,7.53 dB\\
\cline{3-5}\cline{6-8}
\end{tabular}
\end{table}

\begin{table}
\begin{center}
\caption{\textbf{Plane 1 configuration (PC1)\,\,\,\,\,\,\,\,\,\,\,\,\,\,\,\,\,\,\,\,}}
\begin{tabular}{|l|l|l|l|l|l|l|}
\hline
\multicolumn{4}{|l|}{\,\,\,\,\,\,\,\,\,\,\,\,\,\,\,\,\,\,\,\,\,\,\,\,\,\,\,\,\,\,\,\,\,\textbf{Silica}}&\multicolumn{3}{l|}{\,\,\,\,\,\textbf{Chloroform}}\\
\hline
Phase ($\Delta\theta$)&\,\textbf{1.4}&\,\textbf{1.5}&\,\textbf{1.6}&\,\textbf{1.4}&\,\textbf{1.5}&\,\textbf{1.6}\\
\hline
&4.03&2.02&0.40&4.42&2.45&0.88\\
\cline{2-7}
XR1 (dB)&1.49&3.35&1.49&0.39&1.78&0.41\\
\cline{2-7}
&2.72&2.72&2.72&2.72&2.72&2.72\\
\hline
\textbf{FOM}&8.25&8.09&4.62&7.53&6.95&4.02\\
\hline
\end{tabular}
\end{center}
\end{table}

\begin{figure}[tbp]
\begin{center}
\subfigure[]{\label{f4a}
\begin{minipage}
[b]{0.475\textwidth}
\includegraphics[width=1\textwidth]{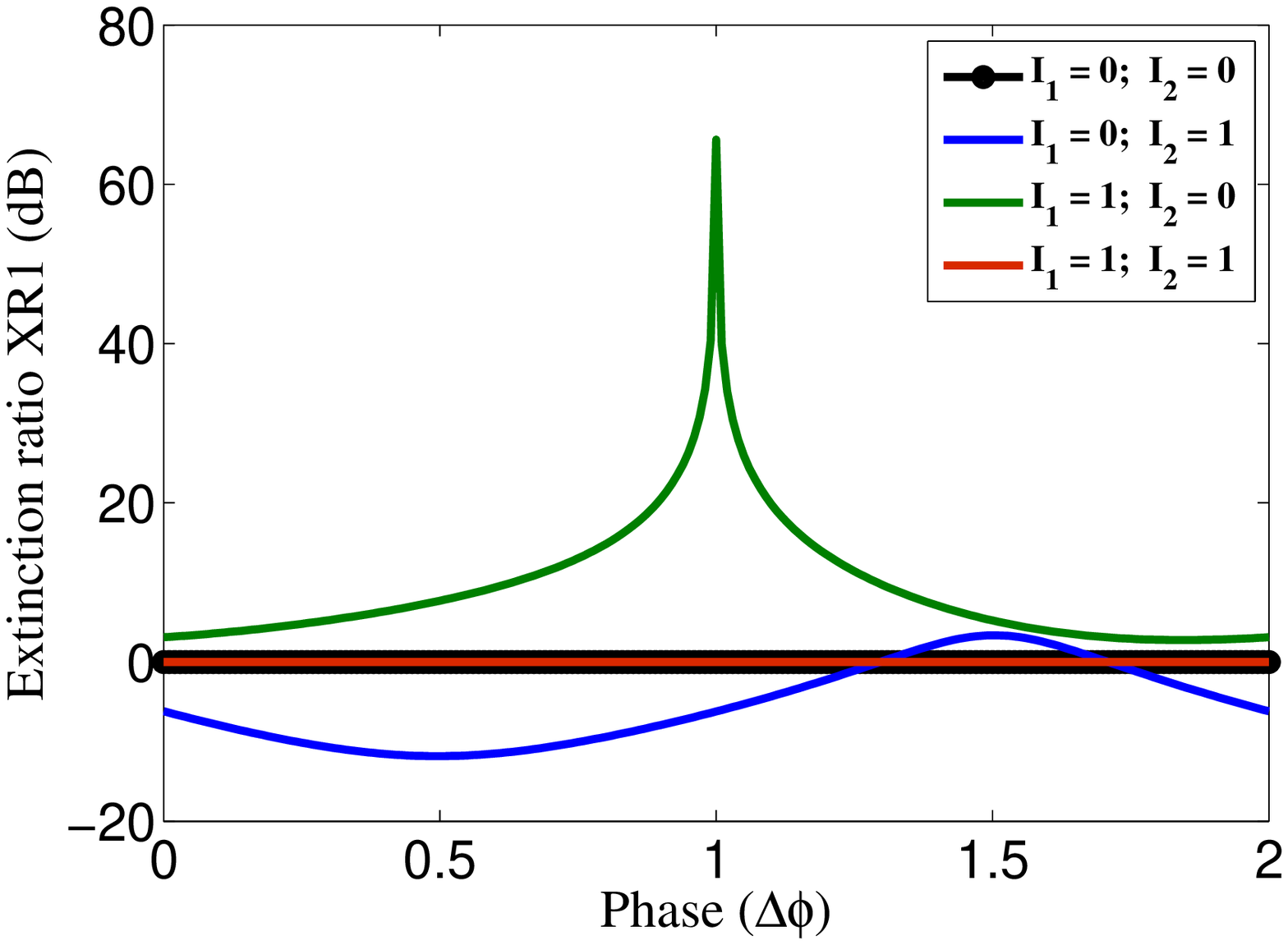}
\end{minipage}}
\subfigure[]{\label{f4b}
\begin{minipage}
[b]{0.475\textwidth}
\includegraphics[width=1\textwidth]{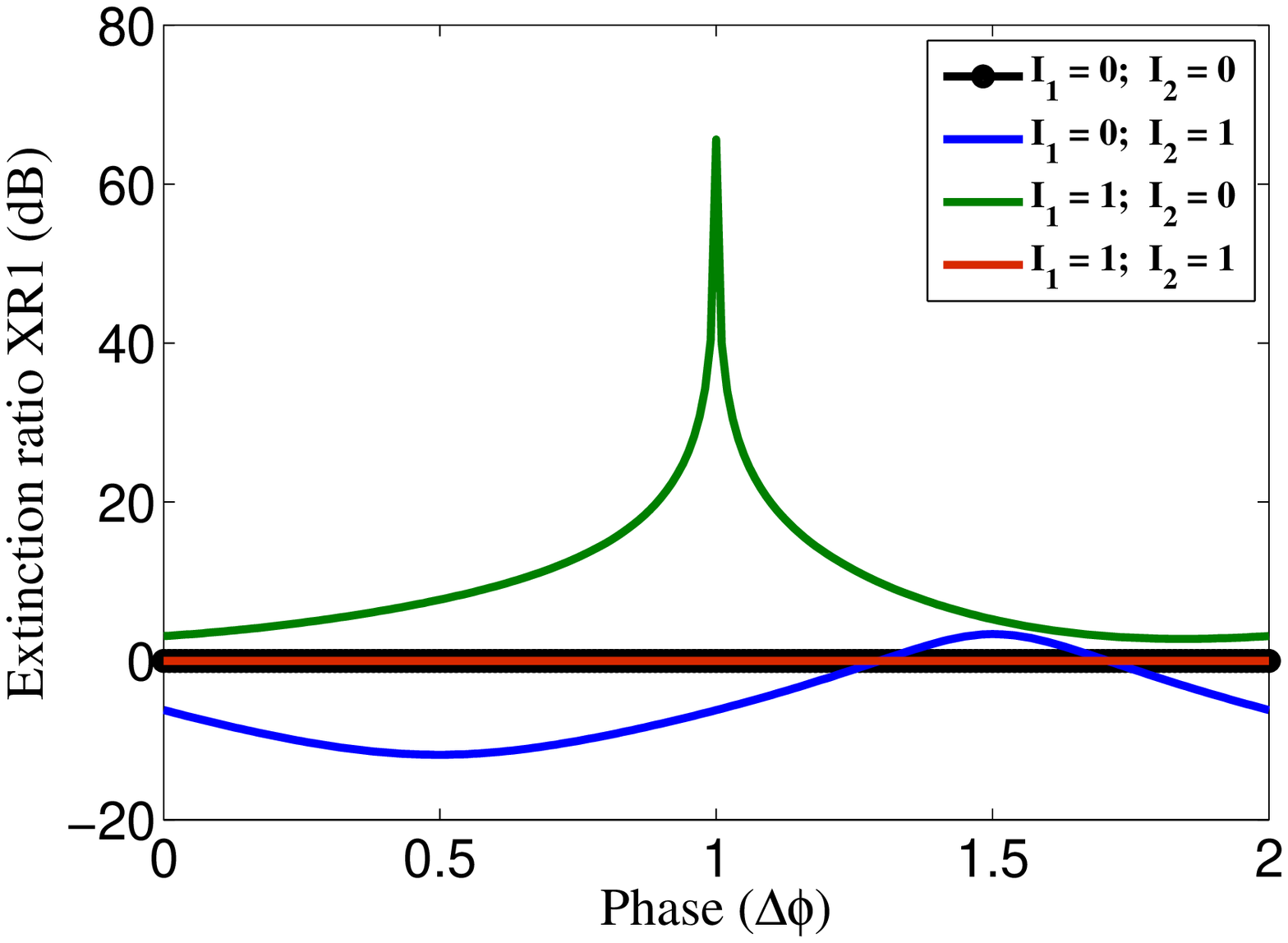}
\end{minipage}}
\end{center}
\caption{ Extinction ratio (XR1) of (a) STPCF and (b) CTPCF of PC2 as a function of the phase parameter  \cite{Uthayakumar}.}
\end{figure}

\begin{table}
\begin{center}
\caption{\textbf{Triangular configuration (TC)\,\,\,\,\,\,\,\,\,\,\,\,\,\,\,\,\,\,\,\,}}
\renewcommand{\arraystretch}{1}
\addtolength{\tabcolsep}{-5pt}
\begin{tabular}{|l|l|l|l|l|l|l|l|}
\hline
\multicolumn{5}{|l|}{\,\,\,\,\,\,\,\,\,\,\,\,\,\,\,\,\,\,\,\,\,\,\,\,\,\,\,\,\,\,\,\,\,\textbf{Silica}\,\,CS\,($\Delta\phi_B$ = $\pi$)}&\multicolumn{3}{l|}{\,\,\,\,\,\,\,\,\,\,\,\,\,\,\,\,\,\,\textbf{Chloroform}\,\,CS\,($\Delta\phi_B$ = $\pi$)}\\
\hline
\textbf{I1}&\textbf{I2}&\,\,\,\,\textbf{CS}&\,\textbf{O1\,(Carry)}\,&\,\textbf{O2\,(Sum)}&\,\,\,\,\textbf{CS}&\,\textbf{O1\,(Carry)}\,&\,\textbf{O2\,(Sum)}\\
\hline
\,\,0&\,\,0&\,\,\,\,\,\,\,0&\,\,\,\,\,\,\,\,\,\,\,\,\,\,\,\,\textbf{0}&\,\,\,\,\,\,\,\,\,\,\,\,\,\textbf{0}&\,\,\,\,\,\,\,0&\,\,\,\,\,\,\,\,\,\,\,\,\,\,\,\,\textbf{0}&\,\,\,\,\,\,\,\,\,\,\,\,\,\textbf{0}\\
\hline
\,\,0&\,\,1&\,\,\,\,\,\,\,1&\,\,\,\,\,-36.64 dB&\,\,\,36.64 dB&\,\,\,\,\,\,\,1&\,\,\,\,\,-36.72  dB&\,\,\,36.72 dB\\
&&&\,\,\,\,\,\,\,\,\,\,\,\,\,\,\,\,\textbf{0}&\,\,\,\,\,\,\,\,\,\,\,\,\,\textbf{1}&&\,\,\,\,\,\,\,\,\,\,\,\,\,\,\,\,\textbf{0}&\,\,\,\,\,\,\,\,\,\,\,\,\,\textbf{1}\\
\hline
\,\,1&\,\,0&\,\,\,\,\,\,\,0&\,\,\,\,\,-12.23 dB&\,\,\,12.23 dB&\,\,\,\,\,\,\,0&\,\,\,\,\,-12.17 dB&\,\,\,12.17 dB\\
&&&\,\,\,\,\,\,\,\,\,\,\,\,\,\,\,\,\textbf{0}&\,\,\,\,\,\,\,\,\,\,\,\,\,\textbf{1}&&\,\,\,\,\,\,\,\,\,\,\,\,\,\,\,\,\textbf{0}&\,\,\,\,\,\,\,\,\,\,\,\,\,\textbf{1}\\
\hline
\,\,1&\,\,1&\,\,\,\,\,\,\,0&\,\,\,\,\,\,\,\,0.00 dB&\,\,\,\,-0.00 dB&\,\,\,\,\,\,\,0&\,\,\,\,\,\,\,\,0.00 dB&\,\,\,\,-0.00 dB\\
&&&\,\,\,\,\,\,\,\,\,\,\,\,\,\,\,\,\textbf{1}&\,\,\,\,\,\,\,\,\,\,\,\,\,\textbf{0}&&\,\,\,\,\,\,\,\,\,\,\,\,\,\,\,\,\textbf{1}&\,\,\,\,\,\,\,\,\,\,\,\,\,\textbf{0}\\
\hline
\multicolumn{2}{l|}{}&\,\textbf{FOM}\,&\,\,\,\,\,\,48.88 dB&\,\,\,\,48.88 dB&\,\textbf{FOM}\,&\,\,\,\,\,\,48.90 dB&\,\,\,\,48.90 dB\\
\cline{3-5}\cline{6-8}
\end{tabular}
\end{center}
\end{table}

\subsection{Plane2 configuration}
For PC2, CS is introduced through core 2 and XR's obtained are displayed in Fig. 4. For the input condition, [I$_1$;I$_2$] = [(0;0),(1;1)], the exit power is
almost equal for O$_1$ and O$_2$ and most of it exits through CS exit.  Next for inputs, [I$_1$;I$_2$] = [(0;1)], the XR plot displays negative value for XR1
indicating that most of the input power is exited via O$_2$ and CS ports. In case of input, [I$_1$;I$_2$] = [(1;0)], the most of the exit power transfers through
O$_1$ and that through CS output O$_2$ is almost negligible. For PC2, only EX-OR logic gate can be constructed, hence, there is no possibility of obtaining of
half adder.

\begin{figure}[htb]
\begin{center}
\includegraphics[height=5.5cm,width=6.5cm]{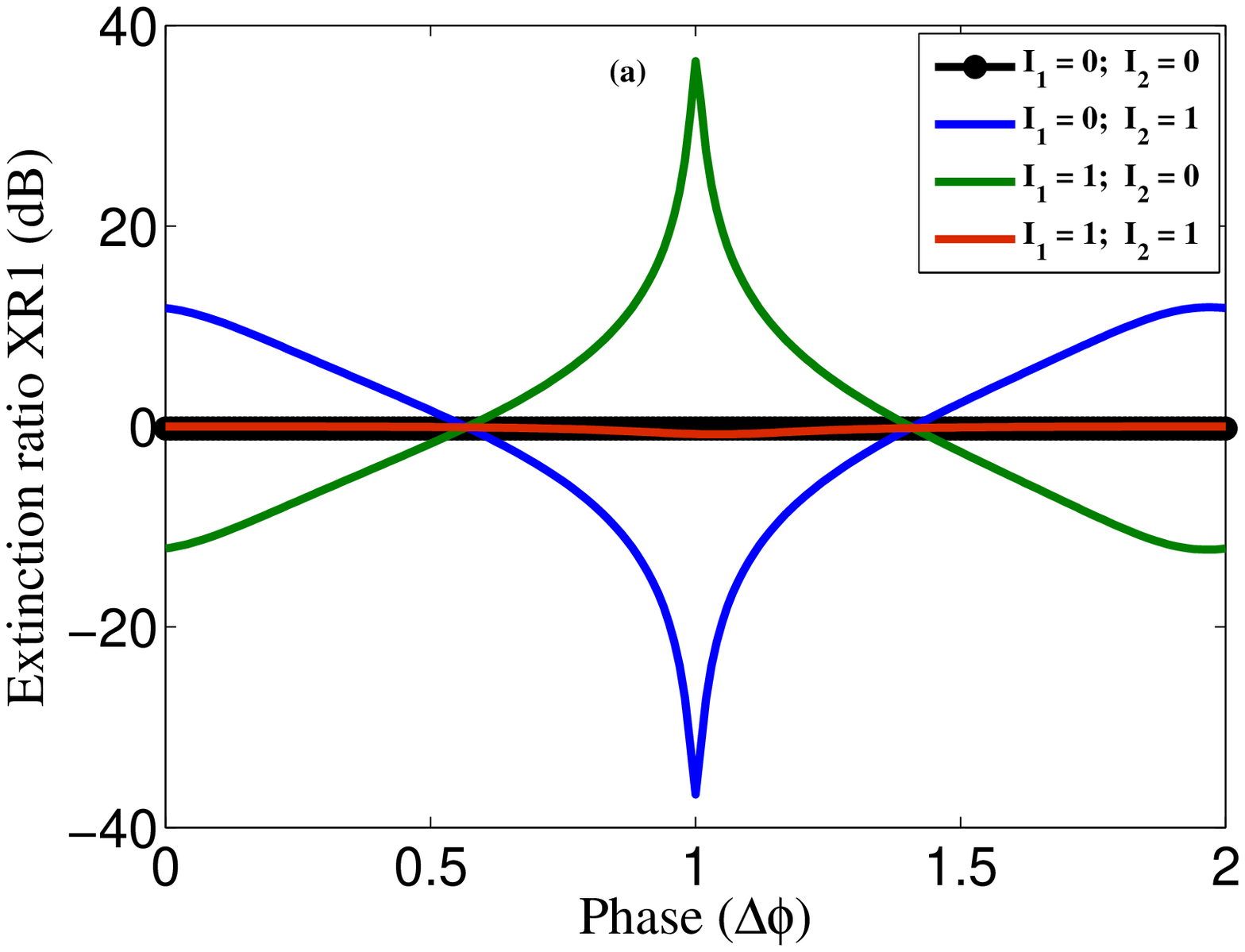}
\includegraphics[height=5.5cm,width=6.5cm]{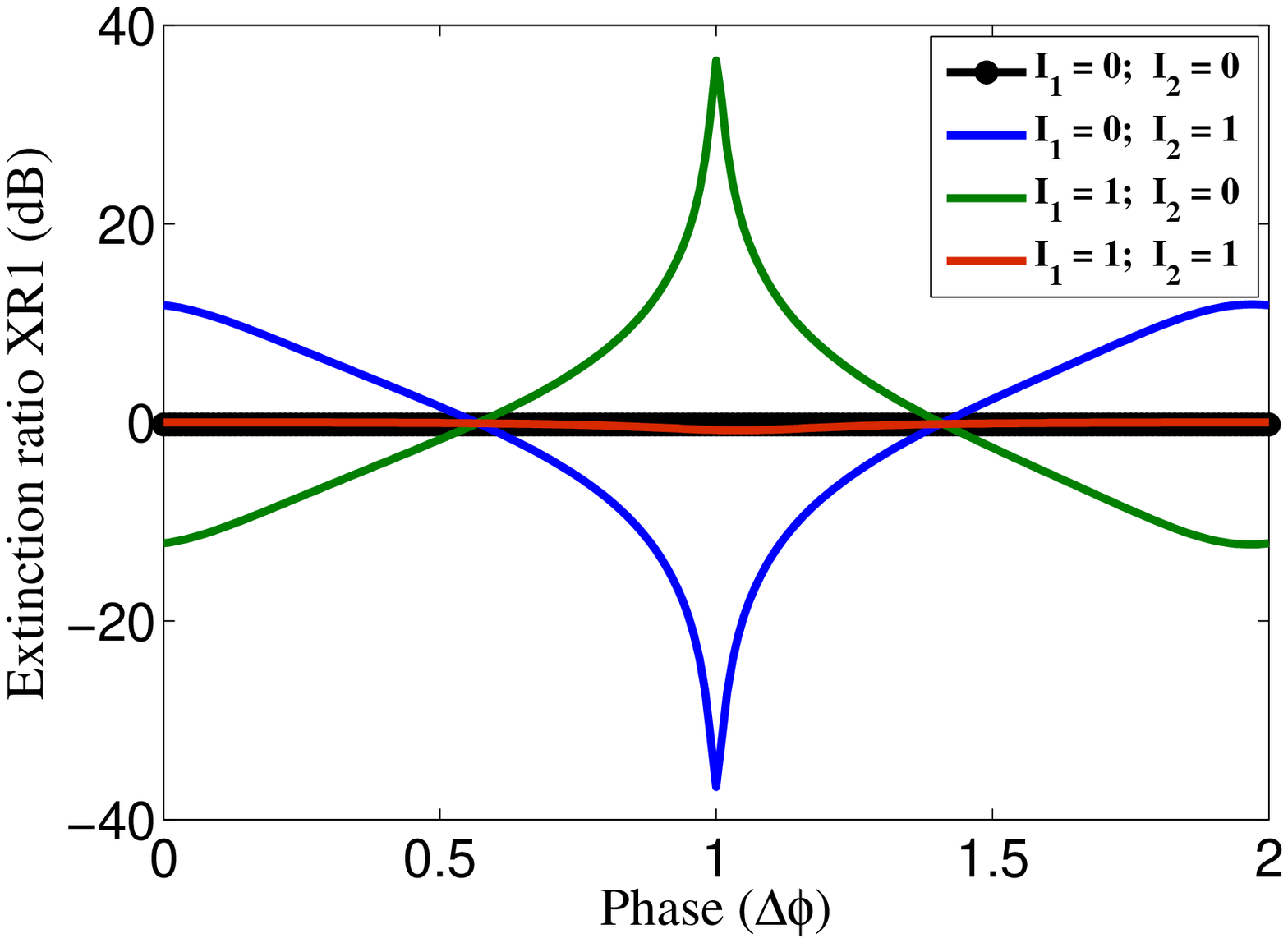}
\end{center}
\caption{Extinction ratio (XR1) of STPCF and CTPCF of TC as a function of the phase parameter.}
\end{figure}

\subsection{Triangular configuration}
XRs for different input combinations for TC is provided in the Fig. 4. For the input condition, [I$_1$;I$_2$] = [(0;0),(1;1)], the XR1 is almost zero, implying
that the exit power through O$_1$ and O$_2$ are almost equal. And most of the exit power is transmitted through CS output. For the condition
[I$_1$;I$_2$] = [(0;1),(1;0)], the input power oscillates between the output ports O$_1$ and O$_2$ with negative and positive values around 36 dB respectively.
Comparing all the configurations considered, TC demonstrates more possibility to construct half adder operation for the multiple phase values
($\Delta\theta=0.6,0.7,0.8,0.9,1,1.1,1.2,1.3,14$). The maximum FOM achieved for the phase value $\Delta\Phi_B=\pi$ is tabulated in Table. 3. And FOM achieved for
other possible phase values are provided in the Table. 4 $\&$ 5 for STPCF and CTPCF.
\begin{table}
\caption{\textbf{Triangular configuration (TC)\,\,\,\,\,\,\,\,\,\,\,\,\,\,\,\,\,\,\,\,}}
\renewcommand{\arraystretch}{1.5}
\addtolength{\tabcolsep}{-6pt}
\begin{center}
\begin{tabular}{|l|l|l|l|l|l|l|l|l|l|}
\hline
\multicolumn{10}{|l|}{\,\,\,\,\,\,\,\,\,\,\,\,\,\,\,\,\,\,\,\,\,\,\,\,\,\,\,\,\,\,\,\,\,\,\,\,\,\,\,\,\,\,\,\,\,\,\,\,\,\,\,\,\,\,\,\,\,\,\,\,\,\,\,\,\,\,\,\,\,\,\,\,\,\,\,\,\,\,\,\,\,\,\textbf{Silica}}\\
\hline
\,Phase ($\Delta\theta$)\,&\,\,\,\,\,\,1\,&\,\,\,\,0.6&\,\,\,\,0.7&\,\,\,\,0.8&\,\,\,\,0.9&\,\,\,\,1.1&\,\,\,\,1.2&\,\,\,\,1.3&\,\,\,\,1.4\\
\hline
&\,36.65\,&\,\,\,0.58&\,\,\,3.53&\,\,\,\,7.36&\,\,13.52\,&\,\,13.60\,&\,\,\,7.40\,&\,\,\,3.56\,&\,\,\,0.61\,\\
\cline{2-10}
\,\,XR1 (dB)&\,12.23\,&\,\,12.23\,\,&\,\,12.23\,\,&\,\,12.23\,\,&\,\,12.23\,\,&\,\,12.23\,\,&\,\,12.23\,\,&\,\,12.23\,\,&\,\,12.23\,\,\\
\cline{2-10}
&\,\,0.00&\,\,\,0.00&\,\,\,0.00&\,\,\,0.00&\,\,\,0.00&\,\,\,0.00&\,\,\,0.00&\,\,\,0.00&\,\,\,0.00\\
\hline
\,\textbf{FOM}&\,48.88\,&\,12.81\,&\,15.76\,&\,19.59\,&\,25.75\,&\,25.83\,&\,19.63\,&\,15.79\,&\,12.84\,
\\
\hline
\end{tabular}
\end{center}
\end{table}

\begin{table}
\begin{center}
\caption{\textbf{Triangular configuration (TC)\,\,\,\,\,\,\,\,\,\,\,\,\,\,\,\,\,\,\,\,}}
\renewcommand{\arraystretch}{1.5}
\addtolength{\tabcolsep}{-6pt}
\begin{tabular}{|l|l|l|l|l|l|l|l|l|l|}
\hline
\multicolumn{10}{|l|}{\,\,\,\,\,\,\,\,\,\,\,\,\,\,\,\,\,\,\,\,\,\,\,\,\,\,\,\,\,\,\,\,\,\,\,\,\,\,\,\,\,\,\,\,\,\,\,\,\,\,\,\,\,\,\,\,\,\,\,\,\,\,\,\,\,\,\,\,\,\,\,\,\,\,\,\textbf{Chloroform}}\\
\hline
Phase ($\Delta\theta$)&\,\,\,\,\,\,1&\,\,\,0.6&\,\,\,0.7&\,\,\,0.8&\,\,\,0.9&\,\,\,1.1&\,\,\,1.2&\,\,\,1.3&\,\,\,1.4\\
\hline
&\,36.72\,&\,\,0.92\,&\,\,3.79&\,\,7.53&\,13.62&\,13.49&\,\,7.20&\,\,3.27&\,\,0.23\\
\cline{2-10}
\,\,XR1 (dB)&\,12.17\,&\,12.17\,&\,12.17\,&\,12.17\,&\,12.17\,&\,12.17\,&\,12.17\,&\,12.17\,&\,12.17\,\\
\cline{2-10}
&\,\,\,0.00\,\,\,&\,\,\,0.00&\,\,\,0.00&\,\,\,0.00\,\,\,&\,\,\,0.00&\,\,\,0.00&\,\,\,0.00\,\,\,&\,\,\,0.00&\,\,\,0.00\\
\hline
\,\textbf{FOM}\,\,&\,48.90\,&\,13.10\,&\,15.96\,&\,19.71\,&\,25.79\,&\,25.66\,&\,19.37\,&\,15.45\,&\,12.40\,\\
\hline
\end{tabular}
\end{center}
\end{table}

\section{Conclusion}
In this article, all optical logic half adder operation is demonstrated via diverse TPCF configurations by employing SSFM. Out of three combinations PC1, PC2 and 
TC considered, PC2 does not exhibit the combination of XOR and AND logic gates to perform the logic half adder function. In the case of PC1, TPCF demonstrates 
the possibility to construct half adder for three different phase values. And that for TC shows the existence of nine different phase values through which 
Boolean operation for half adder can be performed. The prime advantage of the present study is the identical set of TPCF configurations used to demonstrate the 
all-optical logic gates in our previous report reveals the potential to construct all-optical logic half adder operation. When compared to the half adder realized 
through PC1, that obtained through TC demonstrates an excellent FOM. Moreover it is worth to notice that, the FOM of STPCF is little bit greater than CTPCF, 
however CTPCF exhibits excellent transmission characteristics with minimum input power owing to its elevated nonlinearity. In particular, TC CTPCF will be best 
suitable candidate to function as ultra fast all optical half adder.

\section*{Acknowledgment}
RVJ Raja wishes to thank DST fast track programme for providing financial support.

\end{document}